\documentclass{aastex631}

\shorttitle{Introduction to magnetic star-planet interactions}
\shortauthors{Strugarek \& Shkolnik}

\graphicspath{{./}{figures/}}

\usepackage[utf8]{inputenc}
\usepackage{natbib}

\newcommand{\ASadd}[1]{{{#1}}}

\begin{document}

\title{Introduction to magnetic star-planet interactions}
\author[0000-0002-9630-6463]{Antoine Strugarek}
\affiliation{Université Paris Cité, Université Paris-Saclay, CEA, CNRS, AIM, F-91191, Gif-sur-Yvette, France}
\author[0000-0002-7260-5821]{Evgenya Shkolnik}
\affiliation{School of Earth and Space Exploration, Arizona State University, Tempe, AZ 85287, USA}

\begin{abstract}

The interaction between planets and their host stars is governed by the forces of gravity, radiation, and magnetic fields. For planets orbiting their stars at distances of approximately 10 stellar radii or less, these effects are significantly intensified. Such interactions can be investigated through a combination of photometric, spectroscopic, and spectropolarimetric studies spanning wavelengths from X-rays to radio frequencies. When a hot planet resides within the star's sub-Alfv\'enic radius, magnetic star-planet interactions (SPI) become possible, often observable as stellar activity enhancements influenced by the planet's orbital motion rather than stellar rotation alone. Such interactions offer a unique perspective on the atmospheric erosion and magnetospheric characteristics of close-in exoplanets.
The behavior and impacts of these magnetic interactions are highly sensitive to the magnetic fields of both the planet and its host star. This interplay can influence the magnetic activity of both bodies and has implications for the planet's irradiation levels, orbital migration, and the star's rotational dynamics. By employing phase-resolved observational methods on an expanding sample of hot Jupiter (HJ) systems, researchers can now extend these studies to other compact star-planet systems, including smaller planets in the habitable zones of M dwarfs. In such cases, planetary magnetic fields may play an essential role in enabling conditions for surface habitability.
Efforts to comprehend magnetic SPI have led to extensive advancements in theoretical research and computational modeling. These efforts include investigations into the space weather environments of close-in giant exoplanets. Utilizing hydrodynamical (HD) and magnetohydrodynamical (MHD) simulations, researchers aim to provide both qualitative and quantitative descriptions of SPI. These models also facilitate the detection of SPI as a tool for characterizing planetary properties, particularly magnetic field strengths. In this chapter, we first review notable SPI detections before summarizing the current understanding of the underlying physical mechanisms driving SPI.
\end{abstract}


\section{Observations of magnetic star-planet interactions}
\label{sec:obs}
\subsection{Planet-induced stellar activity}

The magnetic fields of exoplanets offer insights into their internal dynamics \citep{Christensen2009,Yadav2017} and provide constraints on atmospheric mass loss \citep{Khodachenko2021,Gronoff2020}. In the solar system, radio emissions resulting from the interaction between planetary magnetospheres and the solar wind have been confirmed from gas giants and the Earth (e.g. \citealt{Zarka2001}). However, no radio emissions from exoplanets have been conclusively detected so far, possibly due to limitations in sensitivity at the higher emission frequencies and/or the entangled stellar radio signature \citep{turner2021,pineda2023,Trigilio2023}. Nevertheless, detections of magnetic SPI have been reported in several systems, indicating that giant exoplanets in short-period orbits can induce activity on their parent stars' photosphere and upper atmosphere. In this way, the magnetic activity of the host star acts as a means to examine the magnetic field of the planet (e.g., \citealt{shkolnik2018}).

The detection of magnetic SPI in HJ systems is facilitated by the proximity of these exoplanets to their parent stars, typically within the Alfv\'en radius ($\leq$10R$_\odot$ or $\leq$0.1 AU for a Sun-like star). At such close distances, the Alfvén speed is faster than the stellar wind speed, enabling direct magnetic interaction with the stellar surface. Should the planet have a strong enough magnetic field, its field may interact with the stellar corona as the planet orbits the star leading to the magnetic reconnection between the planetary and stellar fields and the propagation of Alfv\'en waves, and the creation of electron beams which can influence the base of the stellar corona. (see \S\ref{sec:Models}).

There has been significant progress in data collection and modeling of magnetic star-planet interactions over the past 20 years. The most compelling evidence for magnetic SPI is the presence of enhanced stellar activity in phase with the planet's orbital motion in excess of activity modulated by the stellar rotation period \citep{Cuntz2000}. The first observation of this phenomenon was reported by \citet{Shkolnik2003}, who observed periodic chromospheric activity through Ca II H \& K variability in the hot-Jupiter host star HD 179949, synchronized with the planet's orbital period of 3.092 days \citep{butler2006} rather than the stellar rotation period of 7 days \citep{fares2012}. These observations involved high-resolution spectroscopy with high signal-to-noise ratios acquired over multiple epochs, \ASadd{and the signature of magnetic SPI in the HD 179949 system were also tentatively identified by \citet{acharya2023}}. Similar Ca II H \& K modulations have been observed in other stars with hot Jupiters, including $\upsilon$ And, $\tau$ Boo, and HD 189733, with such signatures present roughly 70\% of the repeated visits \citep{Shkolnik2005,Shkolnik2008,Gurdemir2012}. However, in some epochs, only rotationally modulated spotting is observed, suggesting variations in the stellar magnetic field configuration that result in weaker or no magnetic SPI with the planet's field. Simulations using magnetogram data of varying solar magnetic fields support this explanation \citep{Cranmer2007}. Additionally, long-term observations of the planet-host star HD 189733 using Zeeman-Doppler Imaging (ZDI) have shown structural evolution in the star's large-scale magnetic field \citep{Moutou2007,Fares2010a,Fares2013a,Fares2017a}.

Planet-phased modulations have also been observed at various wavelengths, including broadband optical photometry (e.g. \citealt{2008A&A...482..691W,2009EM&P..105..373P}), X-ray (e.g. \citealt{acharya2023,2015ApJ...811L...2M}) and far-UV spectroscopy (e.g. \citealt{2015ApJ...805...52P}). However, despite the promise of more direct planetary magnetic field measurements through radio observations \citep{ashtari2022}, efforts to detect a magnetic SPI signature or planetary aurora are still ongoing (e.g., \citealt{Kao2018,Vedantham2020,Kavanagh2023}). Part of the challenge might be that the field strengths are too low to be detected at the relatively low frequencies accessible by current radio arrays (e.g., LOFAR, $\sim$10 MHz, \citealt{2013A&A...556A...2V}). And as had to be done at optical wavelengths, long time baseline observations are needed to better characterize stellar radio variability from which a confident, repeated, planet-phased signal can be extracted. 

The strength of the planetary magnetic field for tidally locked planets has been a topic of debate, but scaling laws predict that it primarily depends on the internal heat flux rather than electrical conductivity or rotation speed \citep{Yadav2017}.  Equipped with both planet-induced stellar activity enhancement measurements from \citet{Shkolnik2008} and \citet{Cauley2018} and the large scale field strengths of the stars from spectropolarimetry, \citet{Cauley2019} found that the surface magnetic field strengths for four hot Jupiters range from 20 G to 120 G,  $\sim$10–100 times larger than Jupiter's field, but in agreement with internal heat flux scaling laws,shedding light on dynamo processes in exoplanets. (see \S\ref{sec:energetics}). 

When characterizing habitable zone planets, low-mass stars are preferred targets since the habitable zone is closer to the parent star compared to the Earth-Sun separation. For detecting magnetic SPI in M dwarf systems, the increased stellar magnetic activity presents both challenges and opportunities. While these environments subject planets to heightened stellar wind pressures and radiation, they also enable the detection of magnetic interactions for lower-mass planets in its habitable zone, where planetary magnetic fields may play a crucial role in protecting atmospheres from erosion. 

The proximity to its star makes the detection and study of these planets easier, and at $\approx$ 10 -- 20 R$_{*}$ they probably orbit within the star's Alfv\'en radius. However, low-mass stars are typically more magnetically active than solar-type stars. Therefore, it is crucial to understand how increased magnetic activity affects the potential habitability of planets orbiting close to low-mass stars and what defenses they possess against it. Detecting and characterizing magnetic SPI in M dwarf planetary systems is essential for addressing these concerns yet present a greater challenge of disentangling the planetary signal from the stellar.

\subsection{Suppressed Stellar Angular Momentum Evolution due to magnetic star-planet interactions}
\label{sec:angular_momentum}

The increasing evidence of interactions between stars and planets impacting stellar activity raises questions about the mechanisms involved and their effects on the rotation of stars. For main sequence stars, the magnetized stellar wind acts as a brake on stellar rotation, causing a decrease in the overall rate of stellar activity as the star ages. This relationship, known as the ``age-rotation-activity" relationship, has been extensively studied.

Both tidal and magnetic star-planet interactions can lead to increased stellar rotation through tidal spin-up or reduced efficiency of magnetic braking \citep{Lanza2010,Cohen2011}. There have even been suggestions that increased rotation can be caused by the engulfment of just one other hot Jupiter in the system's past \citep{qureshi2018}. In all cases, the stars would exhibit higher levels of activity than expected for their age.

This suggests that the age-activity relationship may underestimate the true age of stars hosting planets, making ``gyrochronology" potentially unsuitable for these systems (e.g. \citealt{Gallet2020,Benbakoura2019}), and posing challenges for studying exoplanets and their host stars, including models of planet migration and the evolution of planet atmospheres (see \S\ref{sec:torques_and_migration}).

Several observational studies have indeed found that stars hosting giant planets rotate at faster rates than predicted by evolutionary models (e.g. \citealt{2021ApJ...919..138T}). Concomitantly, the spin-up of the star is generally associated with the migration of the hot Jupiter, which has for instance been measured for WASP-12-b \citep{Wong2022}. Additional evidence for stellar spin-up by a close-in giant planet has been observed in hot Jupiter systems CoRoT-2 and HD 189733 \citep{schroter2011,pillitteri2011}. X-ray studies of their M dwarf companions to these relatively active planet hosts showed no X-ray emission, indicating ages of $>$2 billion years \citep{2014A&A...565L...1P}. However, the rotation-age relationship suggests ages of 100--300 million years for CoRoT-2 and 600 million years for HD 189733.

\section{Physics of star-planet magnetic interactions}
\label{sec:Models}

The physics of magnetic star-planet interactions relies on the existence of a differential motion between a conductive or \ASadd{magnetized} body (the planet \ASadd{and its hypothetical magnetosphere}) and the inhomogeneous medium it bathes in (the stellar wind, \citealt{1958ApJ...128..664P}) that is permeated by a large scale magnetic field originating from the host star. We focus here on the main physical processes that apply to planets close enough to their star to orbit in the sub-Aflvénic part of the stellar wind (see \textbf{Chapter 3, Réville \& Linsky 2023}), \textit{i.e.} such that they can be magnetically connected to their host. Subsequent chapters (\textbf{ Chapter 11 --Zhang \& Sciola 2023; Chapter 12 --Luhmann \& Dong 2023; Chapter 13 --Zarka \& Griessmeier 2023; Chapter 14 --Guedel \& Gaidos 2023}) will expand this description to super-Alfvénic regimes as well (see also \citealt{Cohen2015b} for comparisons between sub-Alfvénic and super-Alfvénic interactions). 

\subsection{Orbital environment of close-in exoplanets and magnetic interactions regimes}
\label{sec:interactionregimes}

The environment of a cool star is dynamic. Because of its magnetic activity and of its hot corona, a cool star drives a thermal wind that accelerates as it escapes away from the star (\textbf{Chapter 3, Réville \& Linsky 2023}). The environment of a cool star is also permeated by magnetic fields that originate from dynamo processes within the star \citep{Strugarek2017,Charbonneau2020} and that structure the plasma conditions along the planetary orbit.  

The planet and its hypothetical magnetosphere can be seen as a conducting obstacle in magnetized flow in the rest frame of the planet. Along the orbit, the obstacle sees a velocity ${\bf v}_0(t)$ that is a composition of the orbital motion and of the stellar wind velocity. Likewise, the obstacle sees a magnetic field ${\bf B}_w(t)$ that varies due to the intrinsic magnetic variability of the star and due to the orbital motion. 

Depending on the characteristics of the obstacle (i.e.~the planet), different regimes can occur for the magnetic star-planet interaction \citep{Strugarek2018a,Saur2018b}. In all cases, the obstacle excites Alfvén waves in the ambient medium. The origin of these waves is either the magnetic reconnection sites where the ambient field reconnect with the planetary magnetosphere, or the distortion of the ambient magnetic field lines when they permeate into the conductive, non-magnetized obstacle if it does not possess a magnetosphere. The excited Alfvén waves propagate along the Elsasser characteristics ${\bf c}_A^{\pm} = {\bf v}_0 \pm {\bf v}_{A}$ away from the obstacle, where ${\bf v}_A = {\bf B}_w/\sqrt{\mu_0 \rho_w}$ is the Alfvén speed in the stellar wind\ASadd{, $\mu_0$ the vacuum magnetic permeability, and $\rho_w$ the stellar wind mass density}. In the context of planets in orbit in the sub-Alfvénic part of the stellar wind, most often the Alfvén speed $v_{A}$ also exceeds the flow seen by the obstacle ${\bf v}_0$ and the Elsasser characteristics are actually very close to the magnetic field lines (see also \citealt{Vidotto2010a}). Alfvén waves therefore travel along magnetic field lines, generally towards the host star. They can be mirrored in the low atmosphere of the star due to the strong gradients in Alfvén speed in the atmosphere of the star, in particular close to the transition region where the density sharply rises from the bottom of the corona  to the chromosphere. If the Alfvén waves are fast enough to bounce and travel back to the orbiting body before it significantly progresses along its orbit (or, equivalently, before the field lines permeating the conductive body have significantly slipped through it), the interaction is generally dubbed \textit{unipolar} and often called a \textit{unipolar inductor} \citep{Laine2012a}. On the opposite, if the Alfvén waves are too slow to perform this back-and-forth, the interaction is generally refereed to as \textit{dipolar} (see \citealt{Strugarek2018a}). 

\subsection{Energetics of magnetic star-planet interactions}
\label{sec:energetics}

\begin{figure}[!htpb]
\centering
\includegraphics[width=\linewidth]{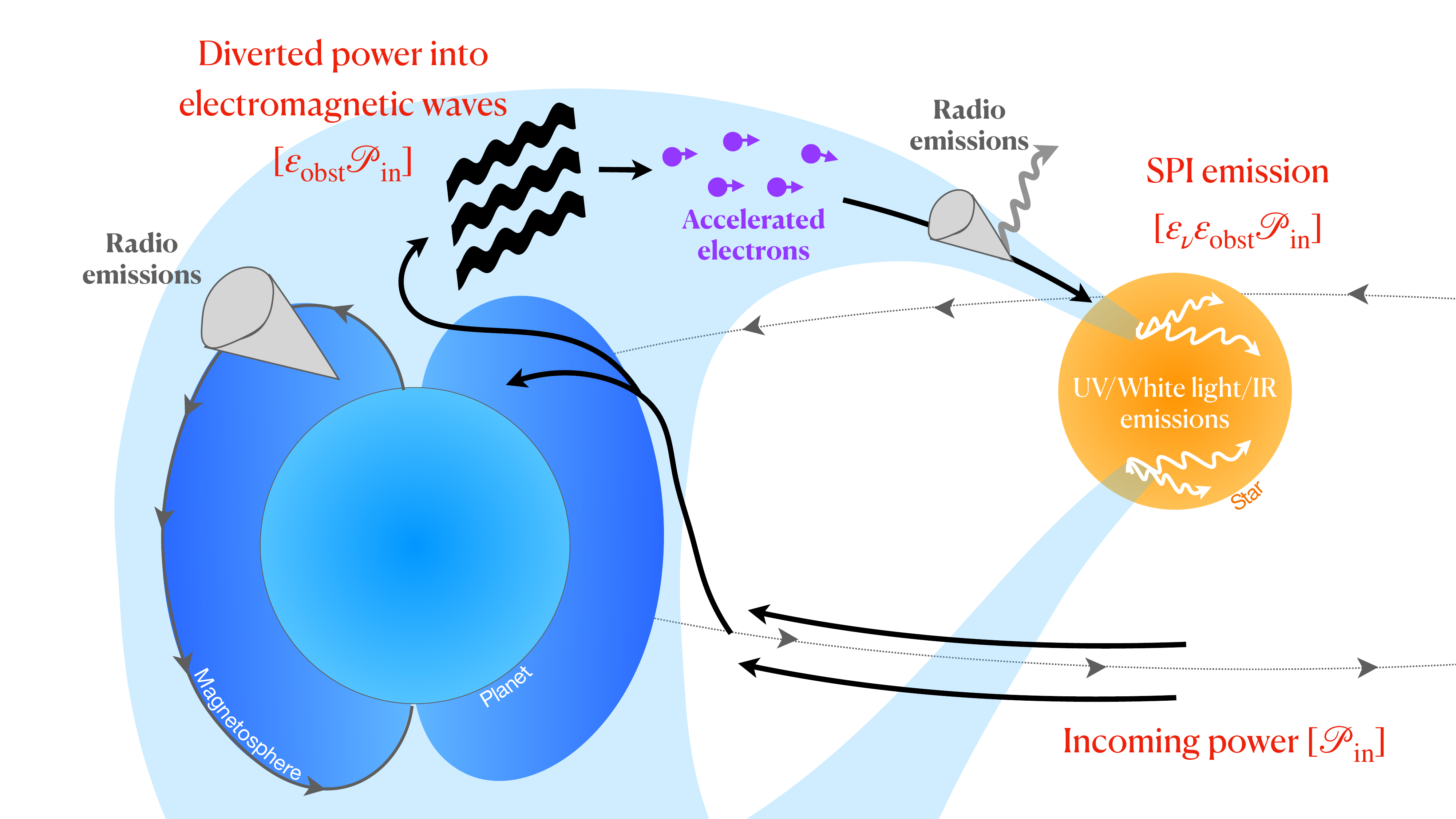}
\caption{Schematic of magnetic star-planet interactions. \label{fig:schematic}}
\end{figure}

After the description of the generic features of SPI (\S \ref{sec:interactionregimes}), we now turn to estimating the energetics associated to star-planet magnetic interactions. The power available to SPI is the Poynting flux ${\bf S}_{\rm in}$ intercepted by the obstacle (planet plus its hypothetical magnetosphere), shown in Figure \ref{fig:schematic}, and which can be generically expressed as 
\begin{equation}
\label{eq:Poynting_in}
{\bf S}_{\rm in} = \frac{\left({\bf v}_0\times{\bf B}_w\right)\times{\bf B}_w}{\mu_0}\, ,
\end{equation}
where ${\bf B}_w$ is the stellar wind magnetic field at the planetary orbit. The power associated to the Poynting flux intercepting the area of the planetary disk can be written as
\begin{equation}
  \label{eq:P_in}
  \mathcal{P}_{\rm in} = 2\pi R_P^2 \left|{\bf S}_{\rm in}\right|\, .
\end{equation}

This Poynting flux intercepts the obstacle, and can in turn be in part channeled away from the planet in the form of Alfvénic perturbations \citep{Neubauer1998}, but it can also power emissions within the planetary magnetosphere that could be detectable in radio \citep{Zarka2001,ashtari2022}. Drawing a parallel with what occurs in the magnetic interaction between Jupiter and its satellites \citep{Saur2013}, it can then be expected that the Alfv\'enic perturbation can deposit part of their energies in the atmosphere of the star. \ASadd{Note, though, that partially ionized outflows from the strongly irradiated upper atmosphere of the planet could modify the interaction scenario borrowed from the Jovian system. Nevertheless, one can expect that the energy carried by the waves will be deposited} either by direct interaction with waves triggered by the star and propagating in the corona and stellar wind, or by the filamentation of the SPI Alfvén waves leading to electron acceleration (like in the atmosphere of Jupiter, see \citealt{2010JGRA..115.6205H}). The net emission power $\mathcal{P}_\nu$ driven by SPI can finally generically be written as 
\begin{equation}
\mathcal{P}_\nu = \varepsilon_{\nu} \varepsilon_{\rm obst} \mathcal{P}_{\rm in}\, ,
\end{equation}
where $\varepsilon_{\rm obst}$ represents the conversion of the Poynting flux seen by the obstacle into the Poynting flux channelled towards the host star, and $\varepsilon_{\nu}$ stands for the conversion factor from the available Poynting flux as Alfvén waves into any kind of emission at a given \ASadd{frequency $\nu$}. It is worth noting that, as of today, these two conversion factors are quantitatively understood relatively well in the context of the magnetosphere of Jupiter, but have not been fully quantitatively assessed in the context of a planet orbiting within a dynamical stellar wind. The first conversion factor, $\varepsilon_{\rm obst}$, has been assessed by various authors in the context of SPI. On one hand, following the Alfvén wing (AW) theory \citep{Neubauer1998}, \citet{Zarka2001} and \citet{Saur2013} estimated this conversion factor to be
\begin{equation}
  \label{eq:FactorAW}
  \varepsilon_{\rm obst}^{AW} = 3 M_a \left( \frac{B_p}{B_w} \right)^{2/3}\, ,
\end{equation}
where $M_a=v_0/v_A$ is the Alfvénic Mach number at the planetary orbit, and $B_p$ is the magnetic field strength at the surface of the planet. Conversely, \citet{Lanza2013a} proposed an alternative mechanism, dubbed \textit{stretch-and-break} (SB), where the saturation of the interaction is considered to occur due to the stretching of magnetic field lines connecting the planet with its environment and which leads to
\begin{equation}
\label{eq:FactorSB}
  \varepsilon_{\rm obst}^{SB} = \left(\frac{B_p}{B_w}\right)^{2} \left(1- \left(1 - \frac{3\left(B_w/B_p\right)^{1/3}}{2+\left(B_w/B_p\right)} \right)^{1/2}\right)\, .
  \end{equation}
Using 3D numerical simulations under the resistive MHD framework, \citet{Strugarek2016c} reproduced the conversion factor from the AW model (Eq. \ref{eq:FactorAW}) and found that the AW scenario was realized before the magnetic field lines connecting the planet with its environment could be stretched enough to activate the SB mechanism. Nevertheless, more realistic models with a more accurate description of magnetic reconnection are likely needed to definitively validate which of the two mechanisms actually dominates in close star-planet systems. Disambiguating the two scenarios is critical to be able to interpret the most recent observations of SPI (see, e.g. \citealt{2023A&A...671A.133E}).  Indeed, the application of both models to the case of HD 189733 \citep{Cauley2019} showed that the detected SPI signal in HD 189733 can be explained only by the higher power level predicted with the SB mechanism. Based on this model, \citet{Cauley2019} were able to estimate the field of four known hot Jupiters to range between 20 and 120 G, assuming $\varepsilon_{\nu}\simeq 0.002$ for the detected chromospheric emission in Ca II H{\&}K bands.

The quantification of $\varepsilon_{\nu}$ nevertheless requires studying in details the interactions between an Alfvén wave originating from the planet and the turbulent stellar wind, the stellar corona, the stellar transition region, and its chromosphere. As of today, the quantitative estimate of star-planet magnetic interactions have assumed $\varepsilon_{\nu}$ between 0.001 and 0.1. Recent work by \citet{2025arXiv250103320P} attempted to quantify from ab initio physics the transmission of Alfv\'en waves through a solar-like transition region, thereby quantify a subpart of $\varepsilon_\nu$ (namely the corona to chromosphere transmission). More theoretical work along those lines is needed to better characterize its value for different kind of observables (namely chromospheric tracer and radio emission, see \S \ref{sec:obs}).

\subsection{Geometrical particularities of emissions associated to star-planet magnetic interactions}

\begin{figure}[!htbp]
\centering
\includegraphics[width=\linewidth]{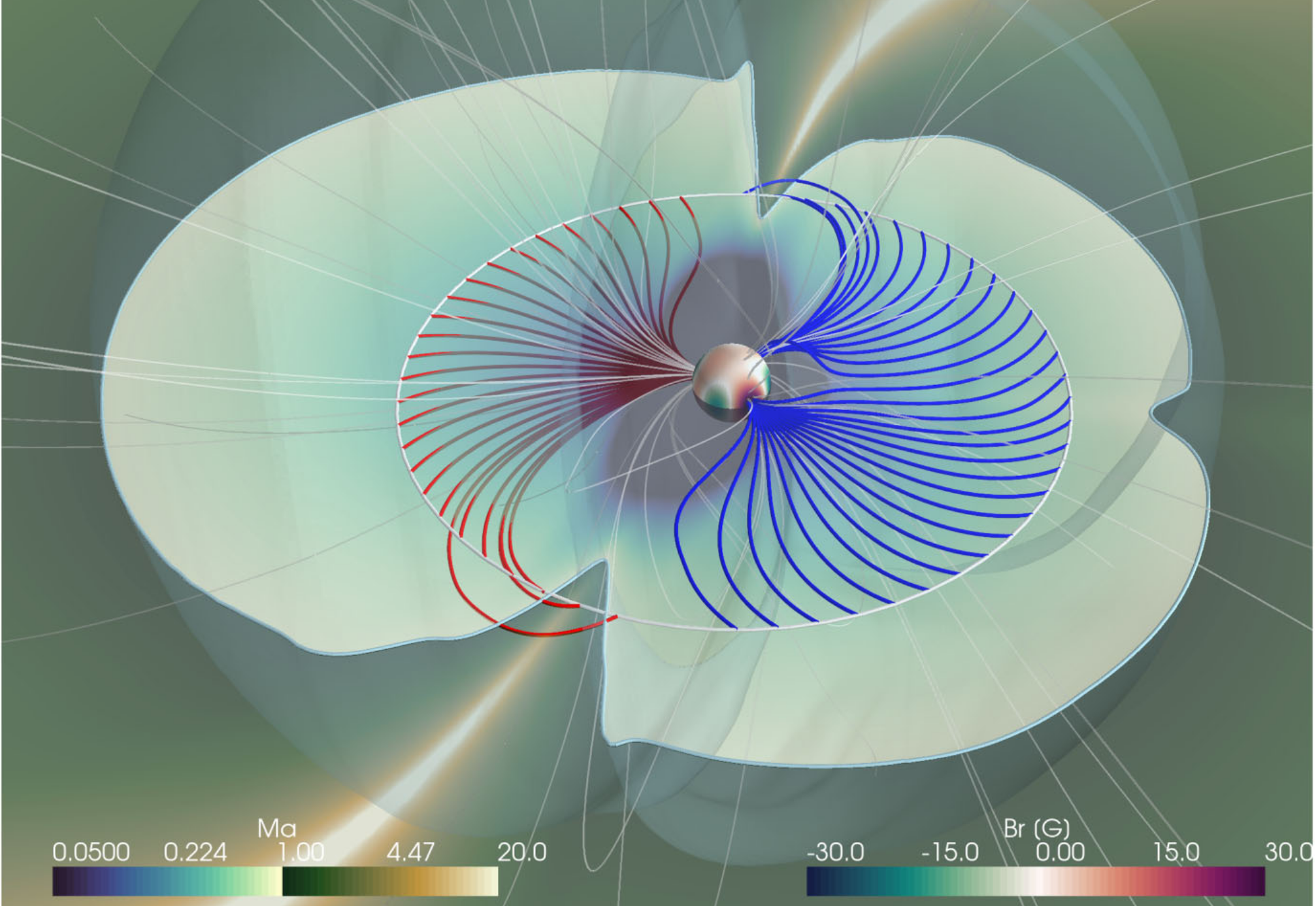}
\caption{Coronal structure of HD 189733 in August 2013, as predicted from the WindPredict stellar wind model \citep{Strugarek2022}. The Alfvénic Mach number is shown on the orbital plane with bright blueish colors ($M_a<1$) and bright yellowish colors ($M_a>1$). The Alfvén 3D surface is shown by the transparent blue surface, and the magnetic field lines by the gray tubes. The magnetic connectivity between the orbit of the planet and the star is shown by the red and blue tubes. \label{fig:HD189733_S22}}
\end{figure}

The theoretical description of SPI is particularly powerful in the sense that any emission triggered by SPI will present a very distinctive temporal modulation. Based on the previous section, the propagation of Alfvén waves close to the planet occurs in the plane defined by $({\bf v}_0,{\bf B}_w)$. These waves then propagate at the local Alfvén speed $v_A$, either into the interplanetary medium or towards the star. The temporality of the energy deposition, and therefore of any SPI-related emission, can therefore be predicted as a function of the orbital phase of the planet provided one can estimate the large-scale magnetic structure of the stellar environment, and the rotation rate of the central star \citep{Ip2004}. This was shown in idealized configuration by \citet{Preusse2006} in 2D, and by \citet{Strugarek2015} in 3D. In the same context, \citet{Hess2011} developed the ExPRES code to predict the temporal modulation of SPI (among other aspects) given a magnetic topology. Leveraging the tomographic reconstruction of the large-scale magnetic field with Zeeman-Doppler Imaging (ZDI), it is now possible to estimate the complex magnetic structuring of the environment of some cool stars and therefore predict in a detailed manner the temporal signature of  SPI signals (e.g. \citealt{Strugarek2019,Kavanagh2021,klein2022}). We show such an example in Figure \ref{fig:HD189733_S22} taken from \citet{Strugarek2022}, where the magnetic connectivity between HD 189733 b and its host star is unveiled as the blue and red tubes connecting the orbital path of the planet (blue circle) to the upper atmosphere of the star. 

\subsection{Torques associated to star-planet magnetic interactions}
\label{sec:torques_and_migration}

In addition to channelling energy in between the two bodies, star-planet magnetic interactions can also lead to changes in the angular momentum budget of a star as well as to exchanges of angular momentum between the orbital motion of the planet and the rotation of the star, as mentioned in Sec.~\ref{sec:angular_momentum}. In the first case, it was theorized by \citet{Lanza2010} that the motion of a magnetized planet in the magnetosphere of a star could significantly affect the large-scale organization of the coronal magnetic field, which in turn significantly changes the net magnetic torque applied to the rotating star (see \textbf{Chapter 3, Réville \& Linsky 2023}). In this work, \citet{Lanza2010} found that stars hosting hot Jupiters could experience significantly slower rates of angular momentum loss compared to similar stars without short-period giant planets, which is consistent with the findings of \citet{Cohen2011} using 3D numerical simulations. Nevertheless, even without invoking such a large-scale effect of the orbiting planet, the planet will always suffer a net torque from its magnetized environment (due to the magnetic tension and the magnetic pressure within the stellar corona), and the angular momentum exchanged in this way will generally transfer to the host star when planets are sufficiently close-in. 

This process was initially considered as well in the context of the magnetosphere of Jupiter, but the magnetic interactions with its natural satellites are not strong enough for this torque to play any role in their orbital motions \citep{Saur2013}. Nevertheless, in the context of star-planet systems, \citet{Strugarek2015} showed that the magnetic torque could lead to migration time-scales down to a few hundred million years, and should therefore be considered when studying the architecture of exo-systems on a secular timescale. This torque was parametrized in \citet{Strugarek2016c}, and systematically compared to the migration torque due to tidal interactions in \citet{Strugarek2017c}. In this latter work, it was generically found that magnetic torques were dominating the migration process for low-mass planets around low mass-stars, and tidal torques were likely dominating for Jupiter-like planets. 

Interestingly, to first order the magnetic and tidal torques always add up to make planets migrate either inward or outward, leading to a spin-up or a spin-down of the host star. Indeed, in the context of tides, the direction of migration is set by the difference between the orbital period and the rotation period of the star. In the context of magnetic interactions, the direction of migration is set by the difference between the orbital period and the rotation period of the corona at the orbit of the planet. For a solar-like star, close to the star the corona is close to being in co-rotation. Henceforth, as long as one considers planets that are sufficiently close to their host, the direction of migration is generally the same for magnetic and tidal interactions.

A self-consistent treatment of the tidal and magnetic torques was more recently carried out by \citet{Ahuir2021a}, who essentially confirmed the findings of \citet{Strugarek2017} and provided for the first time a complete view on the migration of close-in planets after the disk-dissipation phase, following the seminal work of \citet{Zhang2014} that included only tidal interactions. Refining on this approach, \citet{lazovik2023} explored in addition to tidal and magnetic torques the effect of planetary mass-loss in the evolution of close-in planets, providing a quantitative estimate of the probability of existence for hot Jupiters and providing clues to explain the  sub-jovian part of the Neptunian desert \citep{szabo2011}. Finally, more recently, \citet{Garcia2023} showed that magnetic and tidal interactions could explain qualitatively and quantitatively the dearth of hot exoplanets around fast-rotating stars initially identified by \citet{McQuillan2013b}.  

\section{Conclusion}
\label{sec:conclusions}

The detection of magnetic fields of exoplanets allows us to investigate their internal dynamics and gain better insights into the potential atmospheric mass loss caused by stellar wind erosion. So far, the most successful method for detecting magnetic field strengths of exoplanets has been observed signs of interactions between the magnetic fields of the stars and their close-in giant planets. The majority of these searches have focused on main sequence FGK stars, as they are both bright, relatively inactive, and are the dominant hosts of hot Jupiters. They were also the only known hot Jupiter hosts when the first searches began in the early 2000s. FGK stars are advantageous because they exhibit relatively low levels of activity compared to M dwarfs, making it easier to distinguish planetary signals from stellar activity.  

It will be powerful to extend the search to more exoplanet systems with a range of planet masses, so we can understand the range and distribution of possible exoplanet magnetic field strengths. This will also allow us to make quantitative predictions about radio flux densities for stars displaying signs of star-planet interactions, and provide the best targets on which to focus intensive radio observations. As planet searches on the ground and in space discover more close-in planets around M dwarfs, including habitable zone terrestrial planets, the search for magnetic SPI will expand to include these low-mass systems. But since they are intrinsically more active than FGK stars, more observing time will be required to disentangle planet signals from normal stellar activity.  Future studies should prioritize multi-wavelength observations and focus on expanding the sample of known SPI systems to include a broader range of stellar and planetary types. 

In parallel with more extensive and intensive observations, advancements in theories and models of magnetic SPI will deepen our understanding of the enhanced activity observed during specific phases of a planet's orbit. Competing theoretical models for the interaction itself still exist today to predict the absolute strength of SPI (see \S\ref{sec:Models}), and more realistic numerical and analytical works are in order to disentangle them. Although, today, we have only rough estimates of the amount of energy provided by magnetic SPI, they do provide our current best observable tracer. Certainly, questions remain: How much energy goes into local heating, at which altitude in the atmosphere of the star, how much goes into the non-thermal radiation associated to e.g. radio emissions? To tackle these, we next need to study the complex non-linear interaction of Alfv\'en waves triggered by SPI with the dynamical and energetic environment of the star, from the corona down to the chromosphere. With these theoretical developments in hand, we will be able to use the detected signals to put strong constraints on the magnetic field of distant exoplanets. 

As such, by integrating observational data with advanced modeling efforts, we are beginning to disentangle the complex mechanisms driving these phenomena. 



\begin{acknowledgments}
A.S. acknowledges funding from the Programme National de Planétologie (INSU/PNP), from the European Union’s Horizon-2020 research and innovation programme (grant agreement no. 776403 ExoplANETS-A), from the European Research Council project ExoMagnets (grant agreement no. 101125367), and the PLATO/CNES grant at CEA/IRFU/DAp. 
\end{acknowledgments}

{\textbf{Data Availability Statement}} \\
No new data was generated, analyzed, or presented in the manuscript.


\begin{thebibliography}{73}
\expandafter\ifx\csname natexlab\endcsname\relax\def\natexlab#1{#1}\fi

\bibitem[{{\it {Acharya} et~al.\/}(2023){\it {Acharya}, {Kashyap}, {Saar},
  {Singh}, and {Cuntz}\/}}]{acharya2023}
{Acharya}, A., V.~L. {Kashyap}, S.~H. {Saar}, K.~P. {Singh}, and M.~{Cuntz},
  {X-Ray Activity Variations and Coronal Abundances of the Star-Planet
  Interaction Candidate HD 179949}, {\it \apj\/}, {\it 951\/}, 152, 2023.

\bibitem[{{\it Ahuir et~al.\/}(2021){\it Ahuir, Strugarek, Brun, and
  Mathis\/}}]{Ahuir2021a}
Ahuir, J., A.~Strugarek, A.-S. Brun, and S.~Mathis, Magnetic and tidal
  migration of close-in planets, {\it Astronomy \& Astrophysics\/}, {\it
  650\/}, A126, 2021.

\bibitem[{{\it Ashtari et~al.\/}(2022){\it Ashtari, Sciola, Turner, and
  Stevenson\/}}]{ashtari2022}
Ashtari, R., A.~Sciola, J.~D. Turner, and K.~Stevenson, Detecting
  {{Magnetospheric Radio Emission}} from {{Giant Exoplanets}}, {\it The
  Astrophysical Journal\/}, {\it 939\/}, 24, 2022.

\bibitem[{{\it Benbakoura et~al.\/}(2019){\it Benbakoura, R{\'e}ville, Brun,
  {Le Poncin-Lafitte}, and Mathis\/}}]{Benbakoura2019}
Benbakoura, M., V.~R{\'e}ville, A.~S. Brun, C.~{Le Poncin-Lafitte}, and
  S.~Mathis, Evolution of star\textendash planet systems under magnetic braking
  and tidal interaction, {\it Astronomy \& Astrophysics\/}, {\it 621\/}, A124,
  2019.

\bibitem[{{\it Butler et~al.\/}(2006)}]{butler2006}
Butler, R.~P., et~al., Catalog of {{Nearby Exoplanets}}, {\it The Astrophysical
  Journal\/}, {\it 646\/}, 505--522, 2006.

\bibitem[{{\it Cauley et~al.\/}(2018){\it Cauley, Shkolnik, Llama, Bourrier,
  and Moutou\/}}]{Cauley2018}
Cauley, P.~W., E.~L. Shkolnik, J.~Llama, V.~Bourrier, and C.~Moutou, Evidence
  of {{Magnetic Star-Planet Interactions}} in the {{HD}} 189733 {{System}} from
  {{Orbitally Phased Ca II K Variations}}, {\it The Astronomical Journal\/},
  {\it 156\/}, 262, 2018.

\bibitem[{{\it Cauley et~al.\/}(2019){\it Cauley, Shkolnik, Llama, and
  Lanza\/}}]{Cauley2019}
Cauley, P.~W., E.~L. Shkolnik, J.~Llama, and A.~F. Lanza, Magnetic field
  strengths of hot {{Jupiters}} from signals of star\textendash planet
  interactions, {\it Nature Astronomy\/}, {\it 3\/}, 1, 2019.

\bibitem[{{\it Charbonneau\/}(2020)}]{Charbonneau2020}
Charbonneau, P., Dynamo models of the solar cycle, {\it Living Reviews in Solar
  Physics\/}, {\it 17\/}, 4, 2020.

\bibitem[{{\it Christensen et~al.\/}(2009){\it Christensen, Holzwarth, and
  Reiners\/}}]{Christensen2009}
Christensen, U.~R., V.~Holzwarth, and A.~Reiners, Energy flux determines
  magnetic field strength of planets and stars, {\it Nature\/}, {\it 457\/},
  167--169, 2009.

\bibitem[{{\it Cohen et~al.\/}(2011){\it Cohen, Kashyap, Drake, Sokolov,
  Garraffo, and Gombosi\/}}]{Cohen2011}
Cohen, O., V.~L. Kashyap, J.~J. Drake, I.~V. Sokolov, C.~Garraffo, and T.~I.
  Gombosi, The {{Dynamics}} of {{Stellar Coronae Harboring Hot Jupiters}}.
  {{I}}. a {{Time-Dependent Magnetohydrodynamic Simulation}} of the
  {{Interplanetary Environment}} in the {{Hd}} 189733 {{Planetary System}},
  {\it The Astrophysical Journal\/}, {\it 733\/}, 67, 2011.

\bibitem[{{\it Cohen et~al.\/}(2015){\it Cohen, Ma, Drake, Glocer, Garraffo,
  Bell, and Gombosi\/}}]{Cohen2015b}
Cohen, O., Y.~Ma, J.~J. Drake, A.~Glocer, C.~Garraffo, J.~M. Bell, and T.~I.
  Gombosi, The {{Interaction}} of {{Venus-Like}}, {{M-Dwarf Planets}} with the
  {{Stellar Wind}} of {{Their Host Star}}, {\it 806\/}, 41, 2015.

\bibitem[{{\it Cranmer et~al.\/}(2007){\it Cranmer, {van Ballegooijen}, and
  Edgar\/}}]{Cranmer2007}
Cranmer, S.~R., A.~A. {van Ballegooijen}, and R.~J. Edgar, Self-consistent
  {{Coronal Heating}} and {{Solar Wind Acceleration}} from {{Anisotropic
  Magnetohydrodynamic Turbulence}}, {\it The Astrophysical Journal Supplement
  Series\/}, {\it 171\/}, 520--551, 2007.

\bibitem[{{\it {Cuntz} et~al.\/}(2000){\it {Cuntz}, {Saar}, and
  {Musielak}\/}}]{Cuntz2000}
{Cuntz}, M., S.~H. {Saar}, and Z.~E. {Musielak}, {On Stellar Activity
  Enhancement Due to Interactions with Extrasolar Giant Planets}, {\it
  \apjl\/}, {\it 533\/}, L151--L154, 2000.

\bibitem[{{\it {Elekes} and {Saur}\/}(2023)}]{2023A&A...671A.133E}
{Elekes}, F., and J.~{Saur}, {Space environment and magnetospheric Poynting
  fluxes of the exoplanet {\ensuremath{\tau}} Bo{\"o}tis b}, {\it \aap\/}, {\it
  671\/}, A133, 2023.

\bibitem[{{\it Fares et~al.\/}(2013){\it Fares, Moutou, Donati, Catala,
  Shkolnik, Jardine, Cameron, and Deleuil\/}}]{Fares2013a}
Fares, R., C.~Moutou, J.-F.~F. Donati, C.~Catala, E.~L. Shkolnik, M.~M.
  Jardine, A.~C. Cameron, and M.~Deleuil, A small survey of the magnetic fields
  of planet-host stars, {\it Monthly Notices of the Royal Astronomical
  Society\/}, {\it 435\/}, 1451--1462, 2013.

\bibitem[{{\it Fares et~al.\/}(2010)}]{Fares2010a}
Fares, R., et~al., Searching for star-planet interactions within the
  magnetosphere of {{HD}} 189733, {\it Monthly Notices of the Royal
  Astronomical Society\/}, {\it 406\/}, 409--419, 2010.

\bibitem[{{\it Fares et~al.\/}(2012)}]{fares2012}
Fares, R., et~al., Magnetic field, differential rotation and activity of the
  hot-{{Jupiter-hosting}} star {{HD}} 179949: {{Magnetic}} field, {{DR}} and
  activity of {{HD}} 17994, {\it Monthly Notices of the Royal Astronomical
  Society\/}, {\it 423\/}, 1006--1017, 2012.

\bibitem[{{\it Fares et~al.\/}(2017)}]{Fares2017a}
Fares, R., et~al., {{MOVES}} - {{I}}. {{The}} evolving magnetic field of the
  planet-hosting star {{HD189733}}, {\it Monthly Notices of the Royal
  Astronomical Society\/}, {\it 471\/}, 1246--1257, 2017.

\bibitem[{{\it Gallet\/}(2020)}]{Gallet2020}
Gallet, F., {{TATOO}}: {{Tidal-chronology}} standalone tool to estimate the age
  of massive close-in planetary systems, {\it Astronomy \& Astrophysics\/},
  {\it 641\/}, A38, 2020.

\bibitem[{{\it García et~al.\/}(2023)}]{Garcia2023}
García, R.~A., et~al., Stellar spectral-type dependence of the dearth of
  close-in planets around fast-rotating stars: Architecture of kepler confirmed
  single exoplanet systems compared to star-planet evolution models, {\it
  Astronomy \& Astrophysics Letters\/}, {\it submitted\/}, 2023.

\bibitem[{{\it Gronoff et~al.\/}(2020)}]{Gronoff2020}
Gronoff, G., et~al., Atmospheric {{Escape Processes}} and {{Planetary
  Atmospheric Evolution}}, {\it Journal of Geophysical Research: Space
  Physics\/}, {\it 125\/}, 2020.

\bibitem[{{\it Gurdemir et~al.\/}(2012){\it Gurdemir, Redfield, and
  Cuntz\/}}]{Gurdemir2012}
Gurdemir, L., S.~Redfield, and M.~Cuntz, Planet-{{Induced Emission
  Enhancements}} in {{HD}} 179949: {{Results}} from {{McDonald Observations}},
  {\it Publications of the Astronomical Society of Australia\/}, {\it 29\/},
  141--149, 2012.

\bibitem[{{\it Hess and Zarka\/}(2011)}]{Hess2011}
Hess, S. L.~G., and P.~Zarka, Modeling the radio signature of the orbital
  parameters, rotation, and magnetic field of exoplanets, {\it Astronomy \&
  Astrophysics\/}, {\it 531\/}, A29, 2011.

\bibitem[{{\it {Hess} et~al.\/}(2010){\it {Hess}, {Delamere}, {Dols},
  {Bonfond}, and {Swift}\/}}]{2010JGRA..115.6205H}
{Hess}, S.~L.~G., P.~{Delamere}, V.~{Dols}, B.~{Bonfond}, and D.~{Swift},
  {Power transmission and particle acceleration along the Io flux tube}, {\it
  Journal of Geophysical Research (Space Physics)\/}, {\it 115\/}, A06,205,
  2010.

\bibitem[{{\it Ip et~al.\/}(2004){\it Ip, Kopp, and Hu\/}}]{Ip2004}
Ip, W.-H., A.~Kopp, and J.-H. Hu, On the {{Star-Magnetosphere Interaction}} of
  {{Close-in Exoplanets}}, {\it The Astrophysical Journal\/}, {\it 602\/},
  L53--L56, 2004.

\bibitem[{{\it {Kao} et~al.\/}(2018){\it {Kao}, {Hallinan}, {Pineda},
  {Stevenson}, and {Burgasser}\/}}]{Kao2018}
{Kao}, M.~M., G.~{Hallinan}, J.~S. {Pineda}, D.~{Stevenson}, and
  A.~{Burgasser}, {The Strongest Magnetic Fields on the Coolest Brown Dwarfs},
  {\it \apjs\/}, {\it 237\/}, 25, 2018.

\bibitem[{{\it Kavanagh and Vedantham\/}(2023)}]{Kavanagh2023}
Kavanagh, R., and H.~Vedantham, Hunting for exoplanets via magnetic star-planet
  interactions: geometrical considerations for radio emission, {\it Monthly
  Notices of the Royal Astronomical Society\/}, {\it submitted\/}, 2023.

\bibitem[{{\it Kavanagh et~al.\/}(2021){\it Kavanagh, Vidotto, Klein, Jardine,
  Donati, and Fionnag{\'a}in\/}}]{Kavanagh2021}
Kavanagh, R.~D., A.~A. Vidotto, B.~Klein, M.~M. Jardine, J.~F. Donati, and
  D.~O. Fionnag{\'a}in, Planet-induced radio emission from the coronae of {{M}}
  dwarfs: {{The}} case of {{Prox Cen}} and {{AU Mic}}, {\it Monthly Notices of
  the Royal Astronomical Society\/}, {\it 504\/}, 1511--1518, 2021.

\bibitem[{{\it Khodachenko et~al.\/}(2021){\it Khodachenko, Shaikhislamov,
  Lammer, Miroshnichenko, Rumenskikh, Berezutsky, and
  Fossati\/}}]{Khodachenko2021}
Khodachenko, M.~L., I.~F. Shaikhislamov, H.~Lammer, I.~B. Miroshnichenko, M.~S.
  Rumenskikh, A.~G. Berezutsky, and L.~Fossati, The impact of intrinsic
  magnetic field on the absorption signatures of elements probing the upper
  atmosphere of {{HD209458b}}, {\it Monthly Notices of the Royal Astronomical
  Society\/}, {\it 507\/}, 3626--3637, 2021.

\bibitem[{{\it Klein et~al.\/}(2022)}]{klein2022}
Klein, B., et~al., One year of {{AU Mic}} with {{HARPS}} \textendash{} {{II}}.
  {{Stellar}} activity and star\textendash planet interaction, {\it Monthly
  Notices of the Royal Astronomical Society\/}, {\it 512\/}, 5067--5084, 2022.

\bibitem[{{\it Laine and Lin\/}(2012)}]{Laine2012a}
Laine, R.~O., and D.~N.~C. Lin, Interaction of {{Close-in Planets}} with the
  {{Magnetosphere}} of {{Their Host Stars}}. {{II}}. {{Super-Earths}} as
  {{Unipolar Inductors}} and {{Their Orbital Evolution}}, {\it The
  Astrophysical Journal\/}, {\it 745\/}, 2, 2012.

\bibitem[{{\it Lanza\/}(2010)}]{Lanza2010}
Lanza, A.~F., Hot {{Jupiters}} and the evolution of stellar angular momentum,
  {\it Astronomy \& Astrophysics\/}, {\it 512\/}, A77, 2010.

\bibitem[{{\it Lanza\/}(2013)}]{Lanza2013a}
Lanza, A.~F., Star-planet magnetic interaction and evaporation of planetary
  atmospheres, {\it Astronomy and Astrophysics\/}, {\it 557\/}, 31, 2013.

\bibitem[{{\it Lazovik\/}(2023)}]{lazovik2023}
Lazovik, Y.~A., Unravelling the evolution of hot {{Jupiter}} systems under the
  effect of tidal and magnetic interactions and mass-loss, {\it Monthly Notices
  of the Royal Astronomical Society\/}, {\it 520\/}, 3749--3766, 2023.

\bibitem[{{\it {Maggio} et~al.\/}(2015)}]{2015ApJ...811L...2M}
{Maggio}, A., et~al., {Coordinated X-Ray and Optical Observations of
  Star-Planet Interaction in HD 17156}, {\it \apjl\/}, {\it 811\/}, L2, 2015.

\bibitem[{{\it McQuillan et~al.\/}(2013){\it McQuillan, Aigrain, and
  Mazeh\/}}]{McQuillan2013b}
McQuillan, A., S.~Aigrain, and T.~Mazeh, Measuring the rotation period
  distribution of field {{M}} dwarfs with kepler, {\it Monthly Notices of the
  Royal Astronomical Society\/}, {\it 432\/}, 1203--1216, 2013.

\bibitem[{{\it Moutou et~al.\/}(2007)}]{Moutou2007}
Moutou, C., et~al., Spectropolarimetric observations of the transiting
  planetary system of the {{K}} dwarf {{HD}} 189733, {\it Astronomy and
  Astrophysics\/}, {\it 473\/}, 651--660, 2007.

\bibitem[{{\it Neubauer\/}(1998)}]{Neubauer1998}
Neubauer, F.~M., The sub-{{Alfv\'enic}} interaction of the {{Galilean}}
  satellites with the {{Jovian}} magnetosphere, {\it Journal of Geophysical
  Research\/}, {\it 103\/}, 19,843--19,866, 1998.

\bibitem[{{\it {Pagano} et~al.\/}(2009){\it {Pagano}, {Lanza}, {Leto},
  {Messina}, {Barge}, and {Baglin}\/}}]{2009EM&P..105..373P}
{Pagano}, I., A.~F. {Lanza}, G.~{Leto}, S.~{Messina}, P.~{Barge}, and
  A.~{Baglin}, {CoRoT-2a Magnetic Activity: Hints for Possible Star-Planet
  Interaction}, {\it Earth Moon and Planets\/}, {\it 105\/}, 373--378, 2009.

\bibitem[{{\it {Parker}\/}(1958)}]{1958ApJ...128..664P}
{Parker}, E.~N., {Dynamics of the Interplanetary Gas and Magnetic Fields.},
  {\it \apj\/}, {\it 128\/}, 664, 1958.

\bibitem[{{\it {Paul} et~al.\/}(2025){\it {Paul}, {Strugarek}, and
  {R{\'e}ville}\/}}]{2025arXiv250103320P}
{Paul}, A., A.~{Strugarek}, and V.~{R{\'e}ville}, {On stellar hotspots due to
  star-planet magnetic interactions: How much power can actually be transmitted
  to the chromosphere?}, {\it arXiv e-prints\/}, p. arXiv:2501.03320, 2025.

\bibitem[{{\it Pillitteri et~al.\/}(2011){\it Pillitteri, Günther, Wolk,
  Kashyap, and Cohen\/}}]{pillitteri2011}
Pillitteri, I., H.~M. Günther, S.~J. Wolk, V.~L. Kashyap, and O.~Cohen,
  X-{{Ray Activity Phased}} with {{Planet Motion}} in {{Hd}} 189733?, {\it
  741\/}, L18, 2011.

\bibitem[{{\it {Pillitteri} et~al.\/}(2015){\it {Pillitteri}, {Maggio},
  {Micela}, {Sciortino}, {Wolk}, and {Matsakos}\/}}]{2015ApJ...805...52P}
{Pillitteri}, I., A.~{Maggio}, G.~{Micela}, S.~{Sciortino}, S.~J. {Wolk}, and
  T.~{Matsakos}, {FUV Variability of HD 189733. Is the Star Accreting Material
  From Its Hot Jupiter?}, {\it \apj\/}, {\it 805\/}, 52, 2015.

\bibitem[{{\it Pineda and Villadsen\/}(2023)}]{pineda2023}
Pineda, J.~S., and J.~Villadsen, Coherent radio bursts from known {{M-dwarf}}
  planet-host {{YZ Ceti}}, {\it Nature Astronomy\/}, {\it 7\/}, 569--578, 2023.

\bibitem[{{\it {Poppenhaeger} and {Wolk}\/}(2014)}]{2014A&A...565L...1P}
{Poppenhaeger}, K., and S.~J. {Wolk}, {Indications for an influence of hot
  Jupiters on the rotation and activity of their host stars}, {\it \aap\/},
  {\it 565\/}, L1, 2014.

\bibitem[{{\it Preusse et~al.\/}(2006){\it Preusse, Kopp, B{\"u}chner, and
  Motschmann\/}}]{Preusse2006}
Preusse, S., A.~Kopp, J.~B{\"u}chner, and U.~Motschmann, A magnetic
  communication scenario for hot {{Jupiters}}, {\it Astronomy and
  Astrophysics\/}, {\it 460\/}, 317--322, 2006.

\bibitem[{{\it Qureshi et~al.\/}(2018){\it Qureshi, Naoz, and
  Shkolnik\/}}]{qureshi2018}
Qureshi, A., S.~Naoz, and E.~L. Shkolnik, Signature of {{Planetary Mergers}} on
  {{Stellar Spins}}, {\it The Astrophysical Journal\/}, {\it 864\/}, 65, 2018.

\bibitem[{{\it Saur\/}(2018)}]{Saur2018b}
Saur, J., Electromagnetic coupling in star-planet systems, {\it Handbook of
  Exoplanets\/}, pp. 1877--1893, 2018.

\bibitem[{{\it Saur et~al.\/}(2013){\it Saur, Grambusch, Duling, Neubauer, and
  Simon\/}}]{Saur2013}
Saur, J., T.~Grambusch, S.~Duling, F.~M. Neubauer, and S.~Simon, Magnetic
  energy fluxes in sub-{{Alfv\'enic}} planet star and moon planet interactions,
  {\it Astronomy \& Astrophysics\/}, {\it 552\/}, A119, 2013.

\bibitem[{{\it Schr{\"o}ter et~al.\/}(2011){\it Schr{\"o}ter, Czesla, Wolter,
  M{\"u}ller, Huber, and Schmitt\/}}]{schroter2011}
Schr{\"o}ter, S., S.~Czesla, U.~Wolter, H.~M. M{\"u}ller, K.~F. Huber, and
  J.~H. M.~M. Schmitt, The corona and companion of {{CoRoT-2a}}. {{Insights}}
  from {{X-rays}} and optical spectroscopy, {\it Astronomy \& Astrophysics\/},
  {\it 532\/}, A3, 2011.

\bibitem[{{\it Shkolnik et~al.\/}(2003){\it Shkolnik, Walker, and
  Bohlender\/}}]{Shkolnik2003}
Shkolnik, E., G.~A.~H. Walker, and D.~A. Bohlender, Evidence for
  {{Planet}}-induced {{Chromospheric Activity}} on {{HD}} 179949, {\it The
  Astrophysical Journal\/}, {\it 597\/}, 1092--1096, 2003.

\bibitem[{{\it Shkolnik et~al.\/}(2008){\it Shkolnik, Bohlender, Walker, and
  Collier~Cameron\/}}]{Shkolnik2008}
Shkolnik, E., D.~A. Bohlender, G.~A.~H. Walker, and A.~Collier~Cameron, The
  {{On}}/{{Off Nature}} of {{Star}}-{{Planet Interactions}}, {\it The
  Astrophysical Journal\/}, {\it 676\/}, 628--638, 2008.

\bibitem[{{\it {Shkolnik} and {Llama}\/}(2018)}]{shkolnik2018}
{Shkolnik}, E.~L., and J.~{Llama}, {Signatures of Star-Planet Interactions}, in
  {\it Handbook of Exoplanets\/}, edited by H.~J. {Deeg} and J.~A. {Belmonte},
  p.~20, 2018.

\bibitem[{{\it Shkolnik et~al.\/}(2005){\it Shkolnik, Walker, Bohlender, Gu,
  and K{\"u}rster\/}}]{Shkolnik2005}
Shkolnik, E.~L., G.~A.~H. Walker, D.~A. Bohlender, P.~G. Gu, and
  M.~K{\"u}rster, Hot {{Jupiters}} and {{Hot Spots}}: {{The Short-}} and
  {{Long-Term Chromospheric Activity}} on {{Stars}} with {{Giant Planets}},
  {\it The Astrophysical Journal\/}, {\it 622\/}, 1075--1090, 2005.

\bibitem[{{\it Strugarek\/}(2016)}]{Strugarek2016c}
Strugarek, A., Assessing {{Magnetic Torques}} and {{Energy Fluxes}} in
  {{Close-in Star}}\textendash{{Planet Systems}}, {\it The Astrophysical
  Journal\/}, {\it 833\/}, 140, 2016.

\bibitem[{{\it Strugarek\/}(2018)}]{Strugarek2018a}
Strugarek, A., Models of star-planet magnetic interaction, {\it Handbook of
  Exoplanets\/}, pp. 1833--1855, 2018.

\bibitem[{{\it Strugarek et~al.\/}(2015){\it Strugarek, Brun, Matt, and
  R{\'e}ville\/}}]{Strugarek2015}
Strugarek, A., A.~S. Brun, S.~P. Matt, and V.~R{\'e}ville, Magnetic {{Games
  Between}} a {{Planet}} and {{Its Host Star}}: {{The Key Role}} of
  {{Topology}}, {\it The Astrophysical Journal\/}, {\it 815\/}, 111, 2015.

\bibitem[{{\it Strugarek et~al.\/}(2017{\natexlab{a}}){\it Strugarek, Beaudoin,
  Charbonneau, Brun, and {do Nascimento}\/}}]{Strugarek2017}
Strugarek, A., P.~Beaudoin, P.~Charbonneau, A.~S. Brun, and J.-D. {do
  Nascimento}, Reconciling solar and stellar magnetic cycles with nonlinear
  dynamo simulations, {\it Science\/}, {\it 357\/}, 185--187,
  2017{\natexlab{a}}.

\bibitem[{{\it Strugarek et~al.\/}(2017{\natexlab{b}}){\it Strugarek, Bolmont,
  Mathis, Brun, R{\'e}ville, Gallet, and Charbonnel\/}}]{Strugarek2017c}
Strugarek, A., E.~Bolmont, S.~Mathis, A.~S. Brun, V.~R{\'e}ville, F.~Gallet,
  and C.~Charbonnel, The {{Fate}} of {{Close-in Planets}}: {{Tidal}} or
  {{Magnetic Migration}}?, {\it The Astrophysical Journal\/}, {\it 847\/}, L16,
  2017{\natexlab{b}}.

\bibitem[{{\it Strugarek et~al.\/}(2019){\it Strugarek, Brun, Donati, Moutou,
  and R{\'e}ville\/}}]{Strugarek2019}
Strugarek, A., A.~S. Brun, J.-F. Donati, C.~Moutou, and V.~R{\'e}ville, Chasing
  {{Star}}\textendash{{Planet Magnetic Interactions}}: {{The Case}} of
  {{Kepler-78}}, {\it The Astrophysical Journal\/}, {\it 881\/}, 136, 2019.

\bibitem[{{\it Strugarek et~al.\/}(2022)}]{Strugarek2022}
Strugarek, A., et~al., {{MOVES}} \textendash{} {{V}}. {{Modelling}}
  star\textendash planet magnetic interactions of {{HD}} 189733, {\it Monthly
  Notices of the Royal Astronomical Society\/}, {\it 512\/}, 4556--4572, 2022.

\bibitem[{{\it Szab{\'o} and Kiss\/}(2011)}]{szabo2011}
Szab{\'o}, {\relax Gy}.~M., and L.~L. Kiss, A {{SHORT-PERIOD CENSOR OF
  SUB-JUPITER MASS EXOPLANETS WITH LOW DENSITY}}, {\it The Astrophysical
  Journal\/}, {\it 727\/}, L44, 2011.

\bibitem[{{\it {Tejada Arevalo} et~al.\/}(2021){\it {Tejada Arevalo}, {Winn},
  and {Anderson}\/}}]{2021ApJ...919..138T}
{Tejada Arevalo}, R.~A., J.~N. {Winn}, and K.~R. {Anderson}, {Further Evidence
  for Tidal Spin-up of Hot Jupiter Host Stars}, {\it \apj\/}, {\it 919\/}, 138,
  2021.

\bibitem[{{\it {Trigilio} et~al.\/}(2023)}]{Trigilio2023}
{Trigilio}, C., et~al., {Star-Planet Interaction at radio wavelengths in YZ
  Ceti: Inferring planetary magnetic field}, {\it arXiv e-prints\/}, p.
  arXiv:2305.00809, 2023.

\bibitem[{{\it Turner et~al.\/}(2021)}]{turner2021}
Turner, J.~D., et~al., The search for radio emission from the exoplanetary
  systems 55 {{Cancri}}, {$\upsilon$} {{Andromedae}}, and {$\tau$} {{Bo\"otis}}
  using {{LOFAR}} beam-formed observations, {\it Astronomy \& Astrophysics\/},
  {\it 645\/}, A59, 2021.

\bibitem[{{\it {van Haarlem} et~al.\/}(2013)}]{2013A&A...556A...2V}
{van Haarlem}, M.~P., et~al., {LOFAR: The LOw-Frequency ARray}, {\it \aap\/},
  {\it 556\/}, A2, 2013.

\bibitem[{{\it Vedantham et~al.\/}(2020)}]{Vedantham2020}
Vedantham, H.~K., et~al., Coherent radio emission from a quiescent red dwarf
  indicative of star-planet interaction, {\it Nature Astronomy\/}, pp. 1--7,
  2020.

\bibitem[{{\it Vidotto et~al.\/}(2010){\it Vidotto, Opher, {Jatenco-Pereira},
  and Gombosi\/}}]{Vidotto2010a}
Vidotto, A.~A., M.~Opher, V.~{Jatenco-Pereira}, and T.~I. Gombosi, Simulations
  of {{Winds}} of {{Weak-lined T Tauri Stars}}. {{II}}. {{The Effects}} of a
  {{Tilted Magnetosphere}} and {{Planetary Interactions}}, {\it The
  Astrophysical Journal\/}, {\it 720\/}, 1262--1280, 2010.

\bibitem[{{\it {Walker} et~al.\/}(2008)}]{2008A&A...482..691W}
{Walker}, G.~A.~H., et~al., {MOST detects variability on {\ensuremath{\tau}}
  Bootis A possibly induced by its planetary companion}, {\it \aap\/}, {\it
  482\/}, 691--697, 2008.

\bibitem[{{\it Wong et~al.\/}(2022){\it Wong, Shporer, Vissapragada,
  {Greklek-McKeon}, Knutson, Winn, and Benneke\/}}]{Wong2022}
Wong, I., A.~Shporer, S.~Vissapragada, M.~{Greklek-McKeon}, H.~A. Knutson,
  J.~N. Winn, and B.~Benneke, {{TESS Revisits WASP-12}}: {{Updated Orbital
  Decay Rate}} and {{Constraints}} on {{Atmospheric Variability}}, {\it The
  Astronomical Journal\/}, {\it 163\/}, 175, 2022.

\bibitem[{{\it Yadav and Thorngren\/}(2017)}]{Yadav2017}
Yadav, R.~K., and D.~P. Thorngren, Estimating the {{Magnetic Field Strength}}
  in {{Hot Jupiters}}, {\it The Astrophysical Journal\/}, {\it 849\/}, L12,
  2017.

\bibitem[{{\it Zarka et~al.\/}(2001){\it Zarka, Treumann, Ryabov, and
  Ryabov\/}}]{Zarka2001}
Zarka, P., R.~A. Treumann, B.~P. Ryabov, and V.~B. Ryabov,
  Magnetically-{{Driven Planetary Radio Emissions}} and {{Application}} to
  {{Extrasolar Planets}}, {\it Astrophysics and Space Science\/}, {\it 277\/},
  293--300, 2001.

\bibitem[{{\it Zhang and Penev\/}(2014)}]{Zhang2014}
Zhang, M., and K.~Penev, Stars get dizzy after lunch, {\it Astrophysical
  Journal\/}, {\it 787\/}, 131, 2014.

\end{thebibliography}

\end{document}